\documentclass[twocolumn,aps,prb,groupedaddress]{revtex4}
\usepackage{graphicx}
\usepackage{color}
\usepackage{amsmath}
\usepackage{enumitem}
\usepackage{amssymb}
\usepackage{hyperref}
\usepackage{cancel}
\usepackage{ulem}

\usepackage{multirow}

\newcommand{\be}{\begin{equation}}
\newcommand{\ee}{\end{equation}}

\newcommand{\bea}{\begin{eqnarray}}
\newcommand{\eea}{\end{eqnarray}}

\newcommand{\p}{\partial}

\newcommand{\lb}{\left[}
\newcommand{\rb}{\right]}
\newcommand{\lp}{\left(}
\newcommand{\rp}{\right)}

\newcommand{\Tr}{{\rm \, Tr\,}}

\renewcommand{\vec}[1]{{\boldsymbol #1}}

\newcommand{\addZ}[1]{\textcolor{red}{#1}}

\begin{document}
	
\title{	Superconductivity in the vicinity of an isospin-polarized state in a cubic Dirac band}

\author{Zhiyu Dong and Leonid Levitov} 
\affiliation{Massachusetts Institute of Technology, Cambridge, Massachusetts 02139, USA} 
\date{\today}

\begin{abstract}
We present a theory of superconducting pairing originating from soft critical fluctuations near isospin-polarized states in rhombohedral trilayer graphene. Using a symmetry-based approach, we determine possible isospin order types and derive the effective electron-electron interactions mediated by isospin fluctuations. Superconductivity arising due to these interactions has symmetry and order parameter structure that depend in a unique way on the ``mother'' isospin order. This model naturally leads to a superconducting phase adjacent to an isospin-ordering phase transition, which mimics the behavior observed in experiments. The symmetry of the paired state predicted for the isospin order type inferred in experiments matches the observations. These findings support a scenario of superconductivity originating from electron-electron interactions. 
\end{abstract}
\maketitle

The nature of superconductivity (SC) observed in moir\'e graphene\cite{Cao1} has been a topic of intense interest in recent years.\cite{Balents_review_TBG} Initially, the resemblance between phases diagram in moir\'e graphene and cuprates suggested a non-phonon mechanism of pairing\cite{Cao1}, a scenario supported by 
theoretical investigation of various types of interaction-driven SC\cite{Dodaro,Xu, Guinea,CCLiu,Guo, You_SC, Girish}. Meanwhile, it was noted that the phonon-mediated coupling may be strong enough to explain the observed SC\cite{Wu1, Wu2, Peltonen, Lian}. Subsequent experiments painted a more complex picture\cite{Cao2,Dmitri1,Saito2020,Stepanov,XLiu}, initiating an ongoing debate. 

The recently discovered superconductivity in rhombohedral trilayer graphene (RTG)\cite{HZhou2} offers clues that go a long way towards solving this puzzle.
Though the band structures of RTG and moir\'e graphene are completely distinct, SC in RTG appears at ultralow carrier densities and relatively high temperatures.\cite{HZhou2} This is so because the density of states at low doping where SC occurs is dominated by an extremely flat top of the hole band, reaching values 
as high as those in moir\'e bands.\cite{Koshino,FZhang}
Interestingly, the key aspects of the phase diagrams of RTG and moir\'e graphene resemble each other. 
In moir\'e graphene, the SC domes are found amid a cascade of ordered phases.\cite{Cao1} Likewise, the phase diagram of RTG is packed with various ordered phases\cite{HZhou1} with the SC phases located along the lines 
where symmetry-breaking phase transitions into isospin-polarized states occur. 
E.g. one SC phase (called SC1 in \cite{HZhou1}) appears at the transition from the disordered state with four-fold Fermi sea degeneracy to a valley-polarized state with a two-fold degenerate Fermi sea. Another SC phase (SC2) is found at the transition from a two-fold degenerate Fermi sea to a non-degenerate one. 

The RTG has several appealing aspects as a platform to explore the interplay between strongly correlated phases and superconductivity.
Moir\'e graphene hosts a variety of twist-related defects, such as twist-angle disorder\cite{Wilson,Uri}, heterostrain\cite{Mesple,Huder,Cosma,ZBi} and buckling\cite{SZhou,Uchida,Butz}, that can strongly affect the bandstructure in the moir\'e flatbands. In comparison, RTG is a clean system free of this type of defects. In addition, the band dispersion and electron wavefunctions in moir\'e materials are fairly cumbersome, whereas the RTG enjoys a much simpler band structure\cite{Koshino,FZhang}, allowing for analytical modeling.


\begin{figure}
	\includegraphics[width=0.4\textwidth]{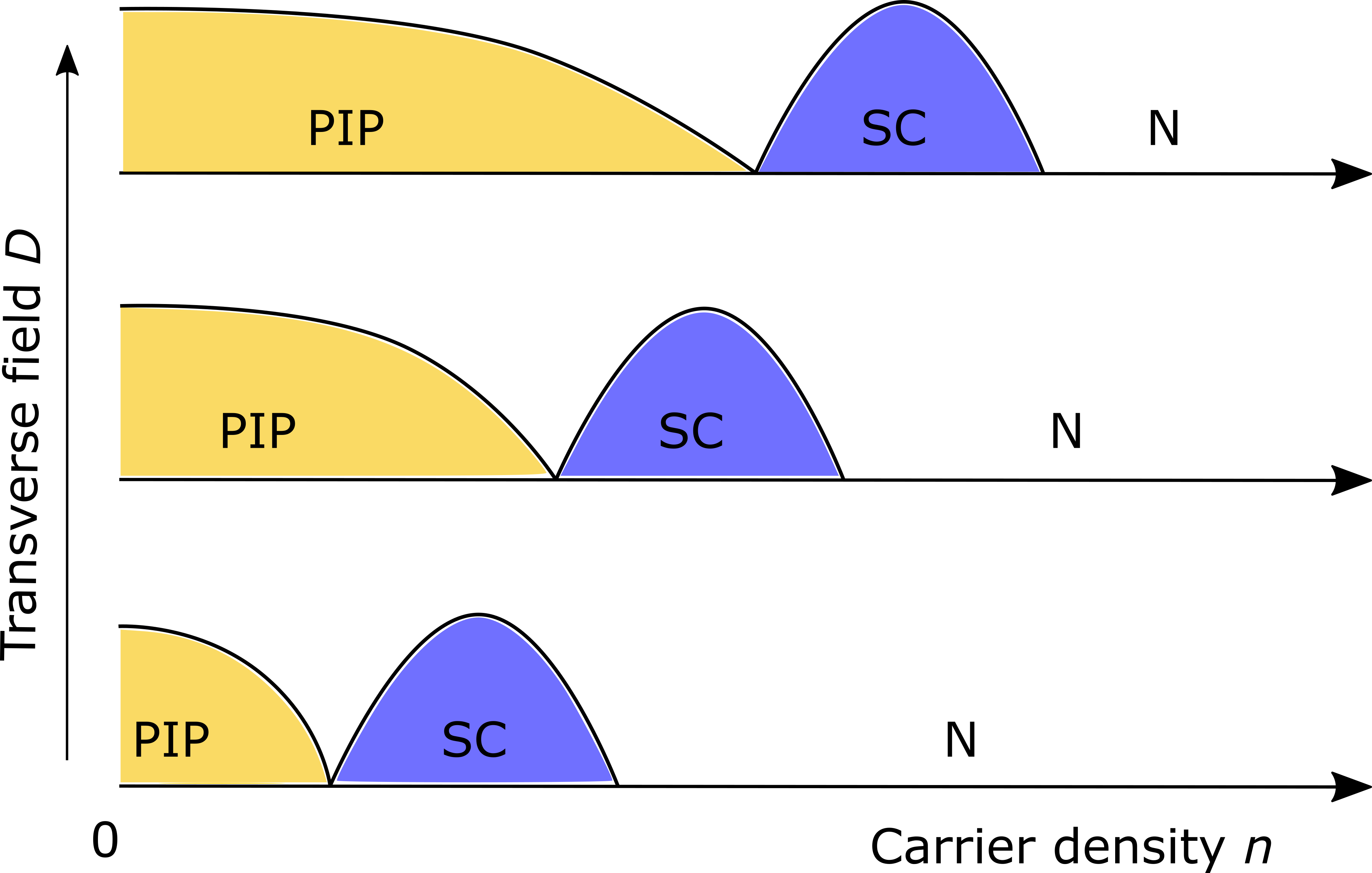}
	\caption{Phase diagram predicted from the model of superconductivity in RTG mediated by critical fluctuations in the partially isospin polarized phase (PIP). The pairing interaction is strongest at the phase boundary between the PIP phase and the unpolarized phase (N), leading to superconductivity (SC) near the PIP phase onset. 
	} \label{fig:phase diagram}
	\vspace{-5mm}
\end{figure}
With this motivation in mind, here we aim at developing 
a framework 
capable of predicting the correct superconducting orders in RTG. 
Accounting for the fact that all SC phases in RTG are located at the vicinity of phase boundaries of isospin-ordered phases, 
we will treat these phases as ``mother states" for the corresponding SC orders, 
and explore the scenario that the critical fluctuations of the order parameter act as a pairing glue\cite{Chubukov_review,Scalapino}. 
Unlike the more conventional 
scenarios\cite{Kopnin2011,Chou} 
our framework  naturally explains the intimate relation between SC and ordered states observed in experiment.
The first step of our analysis is to consider possible orders in the mother state. Taking the SC1 state as an illustration, and using a symmetry-based approach, 
we identify 
a symmetry-breaking phase with the characteristics matching those 
of the valley-polarized spin-unpolarized phase (labeled PIP in Ref.\onlinecite{HZhou2}) near which the SC1 phase is observed. The electron-electron interactions that mediate pairing arise from critical fluctuations, leading to superconductivity near the onset of the order in the partially polarized phase, as illustrated in Fig.\ref{fig:phase diagram}. This framework unambiguously predicts 
a spin-singlet superconductivity, which is compatible with the Pauli-limited SC observed in experiment\cite{HZhou2}. This approach allows a straightforward generalization to describe the interplay between other observed SC phases and their ``mother states.'' 

Although the general idea of fluctuation-mediated pairing is similar to the mechanism studied in iron superconductors,\cite{Hosono_review,Kuroki,Mazin} there are some crucial differences. 
Namely, in iron pnictides, the pairing arises mainly from the pairing-hopping process in which a Cooper pair on one Fermi pocket absorb the momentum of a soft mode and hops to the other Fermi pocket\cite{Kuroki,Mazin}.  In comparison, in our theory for RTG, the Cooper pairs consist of two electrons from opposite valleys $K$ and $K'$. In the process that leads to pairing, two paired electrons exchange valleys as $K, K' \to K',K$, a process mediated by 
a soft mode that carries finite momentum $\pm 2K$. 



\begin{figure}
	\includegraphics[width=0.5\textwidth]{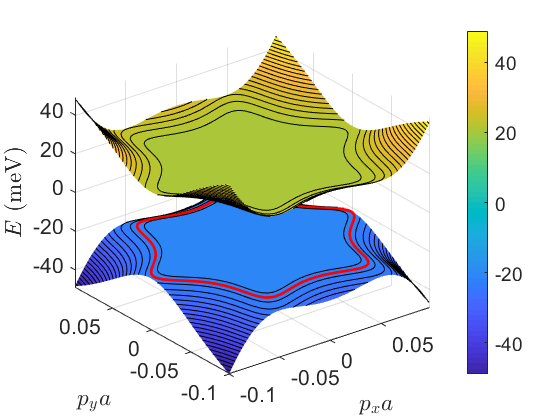}
	\caption{Band dispersion for one valley given by the Hamiltonian in Eq.\eqref{eq:H}, obtained for realistic parameter values: $D = 20$meV, $\alpha/a^3 = 10$eV, $\beta/a^3 = 20$eV, 
	where $a$ denotes the graphene monolayer carbon-carbon spacing $a=1.4\AA$. The red contour marks a Fermi surface at a carrier density comparable to that in experiments\cite{HZhou1,HZhou2}.}\label{fig:band structure}
			\vspace{-5mm}
\end{figure}

We start by writing down the Hamiltonian of electrons in RTG. Restricted by the space-group symmetry, the electron Hamiltonian takes the following form:\cite{Supplement} 
\bea\nonumber
 &&H = \sum_{\vec p} \Psi_{\vec p}^\dagger M (\vec p) \Psi_{\vec p}
,\\ 
 &&M(\vec p) = h_0(\vec p) 1\sigma_1+ h'_0(\vec p) 1_4+h_3(\vec p) 1 \sigma_3 + \nonumber
\\ && + h_1(\vec p)\lp \alpha\tau_3 \sigma_1+ \alpha' \tau_3 1+ \alpha'' \tau_3 \sigma_3 \rp + h_2(\vec p) 1\sigma_2  
\eea
where $\Psi_{\vec p}=\lp \psi_{\vec p,KA},\psi_{\vec p,KB},\psi_{\vec p,K'A},\psi_{\vec p,K'B}\rp^T$, $\tau_i$'s are Pauli matrices in $K$ and $K'$ valley basis, $\sigma_i$'s are Pauli matrices in sublattice (layer) basis [$i=0,1,2,3$, $\sigma_0=\tau_0 = 1_2$]. The quantities $h_n$'s are defined as\cite{Koshino,FZhang}
\bea
&&h_0(\vec p) = \Delta+\frac{p^2}{2 m}, \quad h'_0(\vec p) = \Delta'+\frac{p^2}{2 m'}, \nonumber\\
&&h_1(\vec p)=  p_1^3 - 3p_1p_2^2 , 
\quad h_2(\vec p)=\beta \lp 3p_1^2 p_2 -p_2^3 \rp,\nonumber\\
&&h_3(\vec p) = D +\frac{p^2}{2 m_3}.\label{eq:define h}
\eea
Below, rather than using the realistic values\cite{Koshino,FZhang}, we set $h_0=h_0'=\alpha'=\alpha'' = 0$ and $m_3=\infty$.  As we will see, this choice of parameters represents a simplest case that allows one to
reproduce a qualitatively correct phase diagram
while keeping the analysis simple. For this choice of parameters, the Hamiltonian becomes
\be
H = \sum_{\vec p}\Psi_{\vec p}^\dagger \lb h_1(\vec p) \tau_3\sigma_1 + h_2(\vec p) 1_2\sigma_2 + D 1_2 \sigma_3 \rb \Psi_{\vec p} \label{eq:H}
\ee
The resulting band structure is shown in Fig.\ref{fig:band structure}.
Discarding $h'_0$ and $h'_1$ makes our model Eq.\eqref{eq:H} particle-hole symmetric $E_{p}(\vec p) = - E_{h}(\vec p)$. 
For later convenience, we define 
the absolute value of the energy in the two bands:
\be
E(\vec p) = E_{p}(\vec p) = - E_{h}(\vec p) =\sqrt{h_0(\vec p)^2+h_2(\vec p)^2+D^2}. \label{eq:define_E}
\ee
We note that, while the realistic system is not particle-hole symmetric, the measured RTG phase diagrams on the electron-doped and hole-doped regimes are qualitatively similar --- they both host cascades of ordered states\cite{HZhou1} and superconducting phases\cite{HZhou2}. We therefore proceed with the model in Eq.\ref{eq:H}.

Next, we introduce electron-electron interactions. For the maximal simplicity, we will only consider a local density-density interaction:
\be
H_{\rm int} =V_0 \sum_{\vec p \vec p' \vec q} : \Psi_{\vec p +
\vec q}^{\dagger}1_4 \Psi_{\vec p}  \Psi_{\vec p'-\vec q}^{\dagger}1_4 \Psi_{\vec p'}:
,
\ee
where colons 
indicate normal ordering and the interaction constant $V_0$ is positive, corresponding to a repulsion.

For weakly dispersing $E(\vec p)$ this interaction results in a Stoner instability towards a valley-polarized phase. As an illustration, we consider the limit of a large  transverse field $D$ for which the band can be viewed as 
consisting of two parts: a flat bottom and a steep higher-energy part --- a toy model inspired by the bandstructure in Fig.\ref{fig:band structure}. We will denote the carrier density at which the transition between the two parts occurs as $n_0$, which is a function of 
$D$. In this toy model, we consider only two species: pseudospin up and down ($\uparrow$, $\downarrow$) 
and assume that electrons only interact with electrons of the same species through an exchange interaction $V<0$. Let $\nu_0$ denote the density of states for each species at Fermi surface, and $\nu_1$ and $\nu_2$ be the density of states for each species in the flat and steep parts respectively ($N_1\gg N_2$).  Consider the energy change due to an infinitesimal polarization $n_{\uparrow} = n+\delta n, n_{\downarrow} = n-\delta n$:
\be
\delta E = 2  \delta n^2/\nu_0 + 2 V \delta n^2
\ee
Instability occurs when $\delta E <0$, a condition that gives the Stoner criterion $|V|\nu_0>1$. Thus, so long as $\nu_1<|V|^{-1}<\nu_2$, the instability will occur when $n<n_0(D)$. Given that $n_0$ monotonically increases with $D$, 
the Stoner 
transition 
threshold dependence on $D$ 
is monotonic and approximately linear, as illustrated in Fig.\ref{fig:phase diagram}. 

However, since in RTG there are two valley species and two layer degrees of freedom, the ordering can take a more complicated form, e.g. a valley-only or layer-only polarization. Determining the order in the mother state is crucial, 
since it is the form of the mother state that determines the 
pairing interaction. 

The key question is therefore to identify the orbital channel that has the strongest Stoner instability. To address this question, we 
consider the Stoner criterion for an arbitrary orbital order $\langle \Psi_{\vec p}^\dagger O \Psi_{\vec p}\rangle$:
\be
- V_0\int \frac{d\omega}{2\pi} \int \frac{d^2p}{\lp 2\pi \rp ^2}  G(i\omega,p) O G(i\omega,p)  = O \label{eq:Stoner_matrix}
\ee
where $G(i\omega,p)$ is the electron's Green's function, and $O$ is a general complex-valued $4\times 4$ matrix. 


The multitude of possible ordered states calls for a symmetry-based approach. The system is invariant under the space group P3m1 which is comprised of the point group $C_{3v}$ and translations. 
The three-fold rotation operator $C_3$ leaves the 
Dirac spinors $\Psi$ invariant, whereas the mirror operator $\sigma_d$ swaps $K$ and $K'$ valleys. Translation generates a valley-dependent phase factor when hitting Dirac spinors. From this we can see how Dirac particle-hole bilinears transform under the space group. Namely, the valley-diagonal part of the $O$ matrix contains components that transform under $A_{1}$ and $A_{2}$ representations of $C_{3v}$, and are invariant under lattice 
translations. The valley-off-diagonal part of $O$ matrix contains components that transform under the symmetry group of $K$ point since the total momentum of an intervalley particle-hole pair is $2K$ (or, equivalently $-K$).  
This yields all possible $O_{ij}$ that can be decomposed into three 1D representations  $A_{1,\Gamma}$, $A_{2,\Gamma}$ and $A_{\pm K}$ of the space group. Further classifying them by whether they are even or odd under time reversal (represented by superscripts $+,-$), we find\cite{Supplement}
a list of 
irreducible representations and PIP states with different symmetries given in Table \ref{Table:symmetry_PIP} (a similar analysis can be found in Ref.\onlinecite{Vafek}). Guided by experiment, here we focus on the spin-unpolarized PIP states.



\begin{table}[tbp]
		\begin{tabular}{||c | c ||} 
			\hline
			irrep & Isospin-polarized state orders $O$  \\
			\hline
			$A_{1,\Gamma}^{+}$ & $1_4,1\sigma_1, 1\sigma_3$ \\ 
			\hline
			$A_{1,\Gamma}^{-}$ & $1\sigma_2$ \\ 
			\hline
			$A_{2,\Gamma}^{+}$ & $\tau_3\sigma_2$  \\ 
			\hline
			$A_{2,\Gamma}^{-}$ & $\tau_3 1,\tau_3\sigma_1, \tau_3\sigma_3$\\ 
			\hline
			$A_{\pm K}^+$ & $\tau_1\sigma_n ,\tau_2\sigma_n$, ($n=0,1,3$) \\
			\hline
			$A_{\pm K}^-$ & $\tau_1\sigma_2, \tau_2\sigma_2$\\
			\hline
		\end{tabular}
		\caption{Symmetry classification of different partially isospin-polarized and spin-unpolarized (PIP) states (see text). 
		\vspace{-9mm}
		}
\label{Table:symmetry_PIP}
	\end{table}

To understand which of these orders win over other orders we combine the symmetry analysis with the Stoner criterion, Eq.\ref{eq:Stoner_matrix}. 
On general grounds,\cite{Supplement} the problem of finding the channels with the strongest Stoner instability is equivalent to maximizing the 
quadratic function 
\be
F[O] = -\int \frac{d\omega}{2\pi}\frac{d^2p}{\lp 2\pi \rp ^2}  \frac{1}{4}\Tr \lb O^\dagger G(i\omega,p) O G(i\omega,p)\rb
\ee
subject to the normalization constraint $\frac{1}{4}\Tr O^\dagger O =1$.
Expanding $O$ over the basis of Pauli matrices: $O = d_{ij} \tau_i \sigma_j$, where $d_{ij}$'s are complex valued, we write $F$ into a quadratic form of $d_{ij}$. Maximizing this quadratic form is straightforward since it is diagonal in terms of $d_{ij}$\cite{Supplement}. In this way, we identify the following two channels which have the strongest instability:
\be
O = \tau_1 \sigma_1, \tau_2\sigma_1 
.
\label{eq:mother state order}
\ee
Since the two candidate order parameters transform under 1D representations of the space group, 
their degeneracy is accidental, i.e. not protected by symmetry. 
We therefore conclude that the order parameter will be either $\tau_1\sigma_1$ or $\tau_2\sigma_1$ and not a superposition of the two --- if one of them is condensed, the other will be gapped out. We therefore only need to consider these two possibilities.

Interestingly, both candidate Stoner channels correspond to some valley-coherent order, which matches the observation of the partially isospin polarized (PIP) ordered phase in experiment.\cite{HZhou2} However, which one is ultimately chosen can not be determined unless we introduce a more complicated model, which is beyond the scope of this paper. 
Here we proceed by taking one of them, $O = \tau_1 \sigma_1$, as the 
order parameter of the mother state. 
In the Supplement, we show that 
this choice does not impact our conclusion about superconducting order. 


With this, we can move on to consider the effective e-e interaction generated due to the proximity to this ordered mother state. Similar scenarios have been studied in other systems.\cite{Scalapino,Bickers,Moriya1,Moriya2,Oganesyan,  Roussev, Kim, Chubukov,Chubukov_review,DellAnna,Lederer1,Lederer2} 
Here we employ a scalar-electron coupling model, which is similar to the spin-fermion model studied in literature.\cite{Chubukov,Chubukov_review,Lederer1,Lederer2} 
Namely, we introduce a scalar $\phi$ to represent the order of mother state. The field $\phi$ is softened and fluctuates strongly near the Stoner instability. The Hamiltonian describing the soft mode $\phi$ can be phenomenologically written as 
\be
H_s = \frac{1}{2}|\p_t \phi|^2+ \frac{K}{2}|\p_x \phi|^2 + \frac{J}{2}|\phi|^2  + \frac{U}{2}|\phi|^4,
\ee
where $U$ is always positive, $J>0$ outside $\tau_1 \sigma_1$ phase, $J<0$ inside $\tau_1 \sigma_1$ phase. This Hamiltonian describes a soup of soft-mode fluctuations near the Stoner transition. These coupling between these fluctuations and electrons moving on top of them can be written phenomenologically as 
\be
H_{e-s} = \gamma \sum_{\vec p \vec q}\phi_{\vec q} \psi_{\vec p+\vec q}^\dagger \tau_1 \sigma_1 \psi_{\vec p } +{\rm h.c.}, 
\ee
This expression, which respects the space group symmetry, yields a soft-mode-mediated e-e interaction represented by the following term in system action:\cite{Supplement} 
\be
A_{ee}(\Psi,\overline{\Psi}) = \frac12 \sum_{\omega, \vec p}\tilde{V}_{i\omega,p} \rho^{11}_{i\omega,\vec p}\rho^{11}_{-i\omega,-\vec p}, \label{eq:H_ee1}
\ee
where $\rho^{11}_{i\omega,\vec p} = \sum_{i\nu,\vec k}\overline{\Psi}_{i\nu+i\omega,\vec k+\vec p} \tau_1 \sigma_1 \Psi_{i\nu,\vec k} $, and $\Psi_{i\nu,\vec k}$, $\overline{\Psi}_{i\nu,\vec k}$ are fermionic fields describing Dirac electrons, and
\bea
&&\tilde{V}_{i\omega,p}=
\begin{cases}
	\frac{\gamma^2}{-\omega^2-\omega_{p\parallel}^2}+\frac{\gamma^2}{-\omega^2-\omega_{p\perp}^2}, \quad\text{inside $\tau_1 \sigma_1$ phase},\nonumber\\
	\frac{2\gamma^2}{-\omega^2-\Omega_p^2}, \quad\text{outside $\tau_1 \sigma_1$ phase},\nonumber
\end{cases}
\eea
$\Omega_p = \sqrt{K p^2 + J}$, $\omega_{p\perp} = \sqrt{K} p$, $\omega_{p\parallel} = \sqrt{K p^2 - 2J}$. 
Importantly, the interaction $\tilde{V}(i\omega,p)$ is of a negative sign (an attraction), and is maximized near the phase boundary. In $\tau_1 \sigma_1$ phase, the first term arises from amplitude fluctuation of $\phi$, while the second terms arises from the phase fluctuation of $\phi$. Outside $\tau_1 \sigma_1$ phase, fluctuations along two directions in complex plane contribute equally. 

Below, for 
illustration, we will ignore the momentum and frequency dependence of $\tilde{V}(i\omega,p)$, 
and work with the following simplified effective e-e interaction:
\be
H_{ee} = \sum_{\vec p, \vec k,\vec k'}\tilde{V} 
:\Psi^\dagger_{\vec k'+\vec p} \tau_1 \sigma_1 \Psi_{\vec k'}\Psi^\dagger_{\vec k-\vec p} \tau_1 \sigma_1 \Psi_{\vec k}: 
\label{eq:H_ee2}
\ee
where $\tilde{V}$ is a negative constant estimated above. 

It is important to note that, while the isospin-ordered state acts as a mother state for the SC order through providing pairing glue, the spatial symmetries of these states are in general not directly related. This is so because SC arises outside the isospin-ordered phase --- before the symmetry-breaking ordering occurs. Namely, the soft-mode-mediated pairing interaction respects the full symmetry of the space group. Starting from such a maximally symmetric Hamiltonian, the first symmetry that breaks in SC phase is not necessarily the same as the symmetry of the soft mode that has not condensed yet.

With this in mind, we proceed to develop a symmetry-based analysis of SC order.
Using Eq.\eqref{eq:H_ee2}, the linearized pairing gap equation in an arbitrary pairing channel $\langle\Psi^T\tilde{O}\Psi\rangle$ can be written as
\be
- \int \frac{d\omega}{2\pi} \frac{d^2p}{\lp 2\pi \rp ^2} \tilde{V}  O^T G^T(-i\omega,-p)\tilde{O} G(i\omega,p) O = \tilde{O}, \label{eq:pairing_instability_RTG}
\ee
where $O=\tau_1\sigma_1$ 
is the isospin-polarized state found above and $\tilde{O}$ 
is an arbitrary complex-valued $4\times4$ matrix.

Possible SC orders, classified according their symmetries, are listed in Table\ref{Table:symmetry_SC}. Our goal is to find the one which is most unstable to pairing. Due to the resemblance between Eq.\eqref{eq:pairing_instability_RTG} and  Eq.\eqref{eq:Stoner_matrix}, we will proceed in a similar fashion. First, 
similar to our analysis of Eq.\eqref{eq:Stoner_matrix}, we can convert 
the problem of the most unstable pairing channel to a problem of maximizing 
a quadratic function
\be
\tilde{F}[\tilde{O}] = -\int \frac{d\omega d^2p}{\lp 2\pi \rp ^3}  \frac{1}{4}\Tr \lb \tilde{O}^\dagger O^T G^T(-i\omega,-p) \tilde{O} G(i\omega,p) O\rb \label{eq:tildeF}
\ee
subject to the normalization constraint $\frac{1}{4}\Tr \tilde{O}^\dagger \tilde{O} =1$.

Next, we expand $\tilde{O}$ 
over a Pauli matrix basis $\tilde{O} = \tilde{d}_{ij}\tau_{i}\sigma_{j}$ ($i,j = 0,1,2,3$) where $\tilde{d}_{ij}$'s are complex numbers. Thus, $\tilde{F}$ can be written into a quadratic form of $\tilde{d}_{ij}$. The only off-diagonal terms in this quadratic form are $d_{i3}^*d_{i0}$ and $d_{i0}^*d_{i3}$. 
This fact restricts the candidate channels to take one of the following forms
\be
\tilde{O} = \tau_n \sigma_1, \tau_n\sigma_2, \tau_n (\xi 1 + \eta \sigma_3)
\ee
where $\xi$ and $\eta$ are complex numbers. Evaluating $\tilde{F}$ in the first two cases, and maximizing $\tilde{F}$ by varying $\xi,\eta$ in the third case, we ultimately find that the most unstable pairing channel is
\be
\tilde{O} = \tau_1 (\xi_* 1 +\eta_* \sigma_3). 
\label{eq:strongest_pairing_channel}
\ee
The explicit 
form of $\xi_*$ and $\eta_*$, which is of little relevance for our discussion, is given in the 
Supplement\cite{Supplement}.

The SC order 
in Eq.\eqref{eq:strongest_pairing_channel} corresponds to the irreducible representation $A^{+}_{1,\Gamma}$ which preserves all spatial symmetries. The anticommutation constraint for electrons restricts this pairing channel 
to be a spin-singlet, which is in agreement with the Pauli-limited SC observed in experiment in the SC1 phase.

	\begin{table}[tbp]
		\begin{tabular}{|| c | c | c | c ||} 
			\hline
			\multirow{ 2}{*}{irrep} &  \multicolumn{3}{|c||}{superconducting orders $\tilde{O}$}   \\ [0.5ex] 
			\cline{2-4}
			& spin singlet & spin triplet & spatial symmetry\\
			\hline
			$A_{1,\Gamma}^{+}$  & $\tau_1 1,\tau_1\sigma_1, \tau_1\sigma_3$ &--& \multirow{ 2}{*}{
			no symmetries broken} \\ 
			\cline{1-3}
			$A_{1,\Gamma}^{-}$  &-- & $\tau_1\sigma_2$ &  \\ 
			\hline
			$A_{2,\Gamma}^{+}$  &-- & $\tau_21,\tau_2\sigma_1, \tau_2\sigma_3$ & \multirow{ 2}{*}{reflection symmetry broken}\\ 
			\cline{1-3}
			$A_{2,\Gamma}^{-}$  & $\tau_2\sigma_2$ & -- & \\ 
			\hline
			$A_{\pm K}^+$ & $1_4, 1\sigma_1, 1\sigma_3$ &$\tau_3\sigma_2$  & \multirow{ 2}{*}{pair density wave } \\
			\cline{1-3}
			$A_{\pm K}^-$ & $ \tau_31,\tau_3\sigma_1, \tau_3\sigma_3$ &$1\sigma_2$  & \\
			\hline
		\end{tabular}
		\caption{Superconducting order parameters classified by symmetries.\cite{Sigrist} The SC order found in Eq.\eqref{eq:strongest_pairing_channel} belongs to the $A^{+}_{1,\Gamma}$ representation, which preserves the full space group symmetry. Its spin structure is unambiguously a singlet, in line with experimental observations.  
		}\label{Table:symmetry_SC}
	\end{table}


Before closing, we comment on the unique testable signatures of SC induced by critical fluctuations. The pairing interaction is due to a collective mode with momenta $\sim 2K$ that softens near the isospin-polarization instability. Electrons in different valleys interact by exchanging such soft excitations, generating pair hopping of the form $(K,K')\to(K',K)$,  $(K',K)\to(K,K')$. In the process the momentum of each electron changes by $\pm 2K$, a value large on the scale of $k_F$ and the typical Thomas-Fermi screening parameter. This suggests that introducing screening by a proximal gate\cite{Stepanov,XLiu} 
can serve as a direct test of the pairing mechanism. Indeed, only the relatively long-wavelength harmonics of Coulomb interaction are screened by a gate whereas the 
harmonics responsible for the $\sim 2K$ momentum transfer remain unscreened. Since the long-wavelength harmonics, which are screened out, are responsible for a pair breaking effect in the Cooper channel, we expect SC to be enhanced upon the introduction of screening. An observation of such SC enhancement
would facilitate identifying the unconventional pairing mechanism.

%

In summary, 
superconductivity driven by 
critical soft modes of isospin-polarized states appears to be a viable framework to understand the interplay between different correlated states in RTG, in particular superconductivity observed near the phase boundaries of the isospin-polarized states.\cite{HZhou1, HZhou2} Our symmetry-based approach singles out an isospin-polarized state with a $K$-$K'$ valley coherence which plays a role of a mother state for SC. Accounting for pairing mediated by critical fluctuations in this state 
predicts the correct order in the SC1 phase, in line with experimental observations.\cite{HZhou1, HZhou2}  
This framework is generic and will be straightforward to generalize to other isospin-polarized orders observed in RTG, in particular, the state ``SC2'' that occurs at an interface between so-far-poorly-understood phases with partially and fully polarized isospin. 
A SC phase originating from an isospin-polarized phase with broken time reversal symmetry is expected to be a spin-triplet (and, likely, have a p-wave orbital structure). This is consistent with the observation of SC2 phase being resilient under magnetic fields well in excess of the Pauli threshold fields. Generalizing this approach to describe pairing in other phases is an interesting direction for future work. Once confirmed, it will lend strong support to superconductivity arising exclusively from repulsive interactions.

We thank A. V. Chubukov, V. Cvetkovic, O. Vafek and A. F. Young for useful discussions. After the completion of this work two papers appeared on the arXiv proposing other repulsion-based SC scenarios.\cite{Ghazaryan,Chatterjee}

\pagebreak

\newpage 

\pagebreak

\bigskip

\begin{widetext}
	
	\section{Supplementary Information}
	
	\subsection{Symmetry analysis of rhombohedral trilayer graphene (RTG) with a transverse field}
	
	In this section, we will derive the form of a single-particle Hamiltonian 
	of RTG in a transverse field and discuss its symmetry properties. The symmetry of RTG in the absence of transverse field is $P\overline{3}m1$.\cite{Vafek} Turning on a transverse field breaks the $C_2$ rotation symmetry that swaps the $A$ and $B$ sites in the top and bottom layers. At the same time the mirror symmetries $\sigma_d$ with the mirror planes along the directions of AB bonds remain unbroken. As a result, the space group is lowered to $P3m1$. Then the little group for $\Gamma$ point becomes $C_{3v}$; the little group for $\pm K$ points becomes $C_3$.  The character tables of $C_{3v}$ and $C_3$ are listed below
	\begin{center}
		\begin{table}[htbp]
			\begin{tabular}{|c | c | c | c|} 
				\hline
				$C_{3v}$ &  1 & $2C_3$ & $3\sigma_d$ \\ [0.5ex] 
				\hline\hline
				$A_{1}$ & 1 & 1 & 1   \\ 
				\hline
				$A_{2}$ & 1 & 1 & -1  \\ 
				\hline
				$E$ & 2 & -1 & 0  \\
				\hline
			\end{tabular}
			\qquad
			\begin{tabular}{|c | c | c | c|} 
				\hline
				$C_{3}$ &  1 & $C_3$ & $C_3^{2}$\\ [0.5ex] 
				\hline\hline
				$A$ & 1 & 1 & 1    \\ 
				\hline
				$E_1$ & 1 & $\omega$ & $\omega^* $ \\
				\hline
				$E_2$ & 1 & $\omega^*$ & $\omega$ \\
				\hline
			\end{tabular}
			\caption{Character tables for point groups $C_{3v}$ and $C_{3}$, where $\omega=-\frac12+\frac{\sqrt{3}}{2}i$, $\omega^*=-\frac12-\frac{\sqrt{3}}{2}i$.}
		\end{table}
	\end{center}
	\begin{center}
		\begin{table}[htbp]
			\begin{tabular}{||c | c | c | c||} 
				\hline
				irrep & Dirac bilinears & function of $p$ \\ [0.5ex] 
				\hline\hline
				$A_{1,\Gamma}^{+}$ & $1_4,1\sigma_1,1\sigma_3$ & $1, p^2$   \\ 
				\hline
				$A_{1,\Gamma}^{-}$ & $1\sigma_2$ & $|p|^3 \sin{3\phi_p} = 3p_1^2 p_2 -p_2^3  $  \\ 
				\hline
				$A_{2,\Gamma}^{+}$ & $\tau_3\sigma_2$ & -- \\ 
				\hline
				$A_{2,\Gamma}^{-}$ & $\tau_3 1,\tau_3\sigma_1, \tau_3\sigma_3$ & $|p|^3 \cos{3\phi_p} = p_1^3 - 3p_1p_2^2$  \\ 
				\hline
				$A_{\pm K}^+$ & $\tau_1\sigma_n, \tau_2\sigma_n$, ($n=0,1,3$) &-- \\
				\hline
				$A_{\pm K}^-$  & $\tau_1\sigma_2, \tau_2\sigma_2$ & --\\ 
				\hline
			\end{tabular}
			\caption{The Dirac bilinears and functions of momentum $p$ classified according to the irreducible representations of the space group $P3m1$. 
			\label{tab:symmetry of Dirac bilinear_S} }
		\end{table}
	\end{center}
	
	Below, we determine the form of the single-particle Hamiltonian allowed by symmetry. The Hamiltonian is a bilinear quantity in Dirac spinors $\Psi_{\vec p}=(\psi_{\vec p AK},\psi_{\vec p BK},\psi_{\vec p AK'}, \psi_{\vec p BK'})^T$, defined so that the momentum $\vec p$ for valley $K$ and $K'$ components is measured relative to the corresponding points, $K$ or $K'$. 
	The Hamiltonian in general takes the form of 
	a $4\times 4$ matrix sandwiched between spinors $\Psi_{\vec p}$ and $\Psi_{\vec p}^\dagger$:
			\be
	H = \sum_{\vec p}\Psi_{\vec p}^\dagger M(p)\Psi_{\vec p}, \quad M(p) = h_{ij}(\vec p) \tau_i \sigma_j , \quad i,j = 0,1,2,3. \label{eq:bilinear}
	\ee
As always, the repeated indices are summed over; the quantities $\tau_i$'s are the Pauli matrices in valley basis ($K/K'$), $\sigma_j$'s are the Pauli matrices in sublattice basis ($A/B$). By $\sigma_0=\tau_0=1$ we denote the $2\times2$ identity matrix. 
	
	As in the main text, for conciseness, the spin indices will be suppressed. Hiding spin indices accounts for the fact that our single-particle Hamiltonian is spin-independent and the interactions depend on density but not on spin. The Stoner instability discussed below will involve polarization of valley degrees of freedom and no spin polarization. Likewise, the superconducting orders of interest will have a simple (singlet) spin structure, which justifies using a short-hand notation in which spin indices are suppress, as in Eq.\eqref{eq:bilinear}.
	
	For reader's sake, we comment on how the above expressions would change if the spin indices were included explicitly. 
	The Hamiltonian and spinors would then take the following form  
	\be
	H = \sum_{\vec p,s,s'}\Psi_{\vec p,s}^\dagger M(p)\delta_{ss'}\Psi_{\vec p,s'}, \quad \Psi_{\vec p s}=(\psi_{\vec p AKs},\psi_{\vec p,BKs},\psi_{\vec p AK's}, \psi_{\vec p BK's})^T,  \quad s,s'=\uparrow, \downarrow,
	\ee
where the Kronecker $\delta_{ss'}$	indicates the conservation of spin and the degeneracy of two spin species. This spin structure reflects negligible spin-orbit coupling in graphene-based systems. Therefore, throughout this paper, the short-hand form of Hamiltonian given in Eq.\eqref{eq:bilinear} where all spin indices are suppressed will be sufficient for our needs.
	
	The tensors $\tau_i\sigma_j$ appearing in the Hamiltonian, Eq.\eqref{eq:bilinear}, transform under different irreducible representations of the space group. To determine how these quantities transform, we consider how the elements of the space group act on Dirac spinors. The three-fold rotation operator $C_3$ leaves the 
		states at points $K$ and $K'$
		invariant, whereas the mirror operator $\sigma_d$ 
		swaps the states at points $K$ and $K'$ without causing any other changes. 
	 Translation generates a valley-dependent phase factor when hitting Dirac spinors. Next, we analyze how Dirac particle-hole bilinear quantities transform under the space group. Namely, the valley-diagonal terms transform under either $A_{1}$ or $A_{2}$ representations of $C_{3v}$, and are invariant under lattice translations. The valley-off-diagonal Dirac particle-hole bilinear quantities transform under the symmetry group of $K$ point since the total momentum of an intervalley particle-hole pair is $2K$ (or, equivalently, $-K$).  In this way, we find all possible Dirac particle-hole bilinears can be decomposed into three 1D representations  $A_{1,\Gamma}$ , $A_{2,\Gamma}$ and $A_{\pm K}$ of the space group. Further classifying them by whether they are even or odd under time reversal (represented by superscripts $+,-$), we obtain the second column in Table.\ref{tab:symmetry of Dirac bilinear_S}.
	
	Next, we determine the symmetry of $p$-dependent functions up to cubic form. Basically, $p$-odd and $p$-even terms are even and odd under time reversal. Meanwhile, the cubic term proportional to $\cos 3\phi_p$ (where $\phi_p$ is defined as the angle between $p$ and $K$ vector) picks up a minus sign under the reflection $\sigma_d$ which maps $\phi_p\rightarrow \pi-\phi_p$, so it belongs to $A_{2\Gamma}^-$ representation. In comparison, the cubic term proportional to $\sin 3\phi_p$ is invariant under $\sigma_d$, so it belongs to $A_{1\Gamma}^-$ representation. The results are summarized in the third column of Table.\ref{tab:symmetry of Dirac bilinear_S}. 
	
	
	As the Hamiltonian must be invariant under space group, the momentum dependence of each term in Hamiltonian has to transform in the same way as the Dirac particle-hole bilinear quantities. Therefore, we construct the symmetry-allowed terms by matching the second and third rows in Table\ref{tab:symmetry of Dirac bilinear_S}, yielding the following form of the single-particle Hamiltonian:
	\bea
	&&H = \sum_{\vec p} \Psi_{\vec p}^\dagger M(\vec p)  \Psi_{\vec p}\\
	&&M(\vec p) = h_0(\vec p) 1\sigma_1+ h'_0(\vec p) 1_4+ h_3(\vec p) 1 \sigma_3 + h_1(\vec p)\lp \alpha\tau_3 \sigma_1+ \alpha' \tau_3 1+ \alpha'' \tau_3 \sigma_3 \rp + h_2(\vec p) 1\sigma_2  \nonumber
	\eea
	where
	\bea
	&&h_0(\vec p) = \Delta_0+\frac{p^2}{2 m}, \quad h'_0(p) = \Delta'_0+\frac{p^2}{2 m'}, \\
	&&h_1(\vec p)= \lp p_1^3 - 3p_1p_2^2\rp, \\
	&&h_2(\vec p)=\beta \lp 3p_1^2 p_2 -p_2^3 \rp,\\
	&&h_3(\vec p) = D +\frac{p^2}{2 m_3}.\label{eq:define h_S}
	\eea
These are the expressions used in the main text.

	\subsection{Particle-particle and particle-hole susceptibilities}\label{sec:susceptibility}
	In this section, we derive the form of susceptibilities used in the analysis of instabilities in the particle-hole and particle-particle channels. 
	We start with the 
	Hamiltonian introduced in main text:
	\be
	H = \sum_{\vec p}\Psi_{\vec p}^\dagger \lb h_1(\vec p) \tau_3 \sigma_1  + h_2(\vec p) 1_2\sigma_2 + D 1_2 \sigma_3 \rb \Psi_{\vec p}. \label{eq:H_S}
	\ee
	which gives the electron's Green's function:
	\be
	G(i\omega,\vec p) = \lb i\omega + \mu -  \lp h_1(\vec p)\tau_3 \sigma_1  + h_2(\vec p) 1_2\sigma_2 + D 1_2 \sigma_3 \rp\rb^{-1}
	\ee
	
	On the left hand side of Stoner criterion Eq.
	\eqref{eq:Stoner_matrix}, the integral of two green's functions sandwiching $O$ has the meaning of a particle-hole susceptibility. To solve the Stoner criterion, we have to first evaluate the particle-hole susceptibility. Below, we calculate this quantity in matrix form:
	\bea
	&&\int \frac{d^2 \vec p}{(2 \pi) ^2} \frac{d\omega}{2\pi} G(i\omega,\vec p) \otimes  G(i\omega,\vec p) \nonumber\\
	&&=  \int  \frac{d^2 \vec p}{(2 \pi) ^2} \frac{d\omega}{2\pi} \frac{(i\omega+\mu)^2 1_4 \otimes 1_4  + h_1^2 \tau_3\sigma_1 \otimes \tau_3\sigma_1+ h_2^2 1_2\sigma_2 \otimes1_2\sigma_2+ D^2 1_2 \sigma_3 \otimes 1_2\sigma_3 + (i\omega+\mu) D(1_4\otimes1\sigma_3 + 1\sigma_3\otimes 1_4) }{\lp-(i\omega+\mu)^2+E^2\rp^2} \nonumber\\ 
	&& =   \lb -a_0 1_4 \otimes 1_4  + a_1 \tau_3 \sigma_1 \otimes \tau_3 \sigma_1 + a_2  1_2\sigma_2 \otimes 1_2\sigma_2  + a_3 1_2 \sigma_3 \otimes 1_2\sigma_3 \rb.\label{eq:particlehole_susceptibility_S}
	\eea
Here the $O(D)$ term in the second line vanishes because the quantity under the integral is a full derivative in $\omega$ which vanishes upon integration over $\omega$. Here, we have used that the $h_1$ and $h_2$ are both odd functions of $p$. In the last line, the prefactors $a_0, a_1, a_2, a_3$ are defined as:
	\bea
	&&a_0 = \int \frac{d^2 \vec p}{(2 \pi) ^2} \int  \frac{d\omega}{2\pi} \frac{-(i\omega+\mu)^2}{\lp-(i\omega+\mu)^2+E^2\rp^2} = \int \frac{d^2 \vec p}{(2 \pi) ^2} \frac{1}{4E}\Theta(E-|\mu|) , \\
	&&a_1 = \int \frac{d^2 \vec p}{(2 \pi) ^2} \int  \frac{d\omega}{2\pi} \frac{\alpha^2 \lp p_1^3 - 3p_1p_2^2\rp^2}{\lp-(i\omega+\mu)^2+E^2\rp^2} = \int \frac{d^2 \vec p}{(2 \pi) ^2} \frac{\alpha^2 \lp p_1^3 - 3p_1p_2^2\rp^2}{4E^3} \Theta(E-|\mu|), \\
	&&a_2 = \int \frac{d^2 \vec p}{(2 \pi) ^2} \int  \frac{d\omega}{2\pi} \frac{\beta^2(3p_1^2p_2-p_2^3)^2}{\lp-(i\omega+\mu)^2+E^2\rp^2} = \int \frac{d^2 \vec p}{(2 \pi) ^2} \frac{\beta^2(3p_1^2p_2-p_2^3)^2}{4E^3} \Theta(E-|\mu|), \\
	&&a_3 = \int \frac{d^2 \vec p}{(2 \pi) ^2} \int  \frac{d\omega}{2\pi} \frac{D^2}{\lp-(i\omega+\mu)^2+E^2\rp^2} = \int \frac{d^2 \vec p}{(2 \pi) ^2} \frac{D^2}{4E^3} \Theta(E-|\mu|).
	\eea
	Here, we have obtained the matrix-form particle-hole susceptibility. With this, we can solve the Stoner criterion Eq.
	\eqref{eq:Stoner_matrix} by contracting the quantity evaluated above with the $O$ matrix placed in between the Green's functions $G(i\omega,p)$ and $G(i\omega,p)$.
	
	On the left hand side of the linearized pairing gap equation Eq.
	\eqref{eq:pairing_instability_RTG}, we take only the part that depends on frequency and momentum and carry out the integration, the quantity we get in the particle-particle susceptibility. We evaluate quantity can be evaluate as follows:
	\bea
	&&\int \frac{d^2 \vec p}{(2 \pi) ^2} \int \frac{d\omega}{2\pi} G(i\omega,p) \otimes  G(-i\omega,-p)\nonumber \\
	&&=  \int  \frac{d^2 \vec p}{(2 \pi) ^2} \int \frac{d\omega}{2\pi} \frac{(i\omega+\mu)(-i\omega+\mu)1_4 \otimes 1_4 -  h_1^2 \tau_3 \sigma_1 \otimes \tau_3 \sigma_1 -  h_2^2  1_2\sigma_2 \otimes 1_2\sigma_2 + D^2 1_2 \sigma_3 \otimes 1_2\sigma_3}{\lp-(i\omega+\mu)^2+E^2\rp\lp-(-i\omega+\mu)^2+E^2\rp} \nonumber\\ 
	&& + \int  \frac{d^2 \vec p}{(2 \pi) ^2} \int \frac{d\omega}{2\pi} \frac{
			(i\omega+\mu) D 1_4 \otimes 1_2 \sigma_3 + D (-i\omega+\mu)  1_2 \sigma_3 \otimes 1_4.
		}{\lp-(i\omega+\mu)^2+E^2\rp\lp-(-i\omega+\mu)^2+E^2\rp}\\
	&& =   \tilde{a}_0 1_4 \otimes 1_4  - \tilde{a}_1 \tau_3 \sigma_1 \otimes \tau_3 \sigma_1 - \tilde{a}_2 1_2\sigma_2  \otimes 1_2\sigma_2 + \tilde{a}_3 1_2 \sigma_3 \otimes 1_2\sigma_3 + \tilde{b} \lp 1_4 \otimes 1_2 \sigma_3 + 1_2 \sigma_3 \otimes 1_4 \rp, \label{eq:Cooper_susceptibility_S}
	\eea
	$ \tilde{a}_0, \tilde{a}_1, \tilde{a}_2, \tilde{a}_3, \tilde{b}$ are defined as follows
	\bea
	&&\tilde{a}_0 = \int \frac{d^2 \vec p}{(2 \pi) ^2} \int  \frac{d\omega}{2\pi} \frac{\omega^2+\mu^2}{\lp-(i\omega+\mu)^2+E^2\rp\lp-(-i\omega+\mu)^2+E^2\rp} = \int \frac{d^2 \vec p}{(2 \pi) ^2} \frac{1}{4}\lp \frac{E\Theta(E-|\mu|)}{E^2-\mu^2}  + \frac{|\mu|\Theta(\mu-E)}{\mu^2-E^2} \rp, \nonumber\\
	&&\tilde{a}_1 = \int \frac{d^2 \vec p}{(2 \pi) ^2} \int  \frac{d\omega}{2\pi} \frac{\alpha^2 \lp p_1^3-3p_1p_2^2\rp^2}{\lp-(i\omega+\mu)^2+E^2\rp\lp-(-i\omega+\mu)^2+E^2\rp} = \int \frac{d^2 \vec p}{(2 \pi) ^2} \frac{1}{4}\alpha^2 \lp p_1^3-3p_1p_2^2\rp^2 \lp \frac{\Theta(E-|\mu|)}{E(E^2-\mu^2)} + \frac{\Theta(\mu-E)}{|\mu|(\mu^2-E^2)}\rp, \nonumber\\
	&&\tilde{a}_2 = \int \frac{d^2 \vec p}{(2 \pi) ^2} \int  \frac{d\omega}{2\pi} \frac{\beta^2(3p_1^2p_2-p_2^3)^2}{\lp-(i\omega+\mu)^2+E^2\rp\lp-(-i\omega+\mu)^2+E^2\rp} = \int \frac{d^2 \vec p}{(2 \pi) ^2} \frac{1}{4}\beta^2(3p_1^2p_2-p_2^3)^2 \lp \frac{\Theta(E-|\mu|)}{E(E^2-\mu^2)} + \frac{\Theta(\mu-E)}{|\mu|(\mu^2-E^2)}\rp, \nonumber\\
	&&\tilde{a}_3 = \int \frac{d^2 \vec p}{(2 \pi) ^2} \int  \frac{d\omega}{2\pi} \frac{D^2}{\lp-(i\omega+\mu)^2+E^2\rp\lp-(-i\omega+\mu)^2+E^2\rp} = \int \frac{d^2 \vec p}{(2 \pi) ^2} \frac{1}{4}D^2 \lp \frac{\Theta(E-|\mu|)}{E(E^2-\mu^2)} + \frac{\Theta(\mu-E)}{|\mu|(\mu^2-E^2)}\rp, \nonumber\\
	&&\tilde{b} = \int \frac{d^2 \vec p}{(2 \pi) ^2} \int  \frac{d\omega}{2\pi} \frac{\mu D}{\lp-(i\omega+\mu)^2+E^2\rp\lp-(-i\omega+\mu)^2+E^2\rp} = \int \frac{d^2 \vec p}{(2 \pi) ^2} \frac{1}{4}\mu D \lp \frac{\Theta(E-|\mu|)}{E(E^2-\mu^2)} + \frac{\Theta(\mu-E)}{|\mu|(\mu^2-E^2)}\rp. \nonumber
	\eea 
	$E(\vec p)$ is the absolute value of the energy in the two bands: $E(\vec p)=\sqrt{h_0(\vec p)^2+h_2(\vec p)^2+D^2}$.
	
	Here, we have obtained the matrix-form particle-particle susceptibility. With this, we are able to solve the  linearized pairing gap equation Eq.\eqref{eq:pairing_instability_RTG} by contracting the quantity found above with the $O$ matrix placed in between the Green's functions $G(i\omega,p)$ and $G(i\omega,p)$.
	
	
	\subsection{Stoner instability}
	
	In this section, we derive the 
	variational form of the Stoner criterion used in main text and use it to determine the channel with strongest instability. We start with the matrix form of Stoner criterion for an arbitrary orbital order $\langle \Psi_{\vec p}^\dagger O \Psi_{\vec p}\rangle$:
	\be
	- V_0\int \frac{d\omega}{2\pi} \int \frac{d^2\vec p}{\lp 2\pi \rp ^2}  G(i\omega,\vec p) O G(i\omega,\vec p)  = O, \label{eq:Stoner_matrix_1S}
	\ee
	where $G(i\omega,p)$ is the electron's Green's function and $O$ can be an arbitrary $4\times4$ complex-valued matrix. Since this equation is invariant under multiplying $O$ by a prefactor, we can choose the normalization of $O$ so that $\frac{1}{4}\Tr O^\dagger O =1$. 
	
	Eq. \eqref{eq:Stoner_matrix_1S} is essentially an eigenvalue equation for $O$. To see this, we can express $O$ using Pauli matrices: $O = d_{ij} \tau_i \sigma_j$, where $d_{ij}$ is a 16-dimensional complex-valued vector, dummy indices are summed over. The normalization of $O$ yields the normalization of $d$ vector: $|d_{ij}|^2 =1$. Then, we can define a matrix $M$ so that
	\be
	-\int \frac{d\omega}{2\pi} \int \frac{d^2p}{\lp 2\pi \rp ^2}  G(i\omega,p) \tau_i\sigma_j G(i\omega,p) = M_{ij,i'j'}\tau_{i'}\sigma_{j'}, \label{eq:Stoner_eigenvalue}
	\ee
	then Eq.\eqref{eq:Stoner_matrix_1S} is equivalent to
	\be
	M_{ij,i'j'}d_{i'j'} =  \frac{1}{V_0} d_{ij} ,
	\ee
	which is an eigenvalue equation with $\frac{1}{V_0}$ taking on a role of an eigenvalue.
	
	Our goal is to identify the channel which is most unstable to valley polarization. 
	This is equivalent to finding the channel in which pairing can be triggered by the smallest $V_0$ at zero temperature. Eq.\eqref{eq:Stoner_eigenvalue} indicates that 
	this is equivalent to determining an eigenvector $|d\rangle$ that corresponds to the largest eigenvalue of $M$. Let $\lambda_*$ be the largest eigenvalue, then
	\be
	\lambda_* = \max F[O],\quad F[O]=  \frac{\langle d| M |d \rangle }{\langle d|d \rangle } = -\int \frac{d\omega}{2\pi}\frac{d^2\vec p}{\lp 2\pi \rp ^2}  \frac{1}{4}\Tr \lb O^\dagger G(i\omega,p) O G(i\omega,p)\rb.
	\ee
	where we have used the normalization $\langle d|d\rangle=|d_{ij}|^2=1$. Thus, our problem is equivalent to finding a matrix $O$ (or, equivalently, a vector $|d\rangle$) that maximizes the function $F(O)$.
	
	Next, from Eq.\eqref{eq:particlehole_susceptibility_S}, we notice that the off-diagonal terms vanish upon integration over $\omega$ and $p$. Therefore, 
	$\int d\omega d^2p G(i\omega,p)\otimes  G(i\omega,p)$ only contains the diagonal terms $1_4 \otimes 1_4$, $\tau_3 \sigma_1 \otimes \tau_3 \sigma_1$, $1_2\sigma_2 \otimes 1_2\sigma_2$, $1_2 \sigma_3 \otimes 1_2\sigma_3$. These terms, when sandwiching an arbitrary component 
	of $O$ on the left-hand-side of Eq.\eqref{eq:Stoner_matrix_1S}, can only yield the same matrix component. Say, plugging $O=\tau_i\sigma_j$ in Eq.\eqref{eq:Stoner_matrix_1S} yields an expression proportional to $\tau_i\sigma_j$, and so on. 
	As a result, the function $F(O)$ 
	takes a diagonal form:
	\be
	F[O]=   \sum_{ij}|d_{ij}|^2 \chi_{ij}, \quad \chi_{ij} = -\int \frac{d\omega}{2\pi}\frac{d^2p}{\lp 2\pi \rp ^2}  \frac{1}{4}\Tr \lb \tau_i\sigma_j G(i\omega,p) \tau_i\sigma_j G(i\omega,p)\rb. \label{eq:chi_ij}
	\ee
	The quantity $\chi_{ij}$ has the meaning of a particle-hole susceptibility in the channel $O_{ij}$. From Eq.\eqref{eq:chi_ij} we see that, all that needs to be done to maximize $F$ is to find the largest susceptibility $\chi_{ij}$ and set $d_{ij}=1$, letting other components $d_{i'j'}$ vanish. 
	Accordingly, we plug Eq.\eqref{eq:particlehole_susceptibility_S} into Eq.\eqref{eq:Stoner_matrix_1S} to obtain 
	\be
	\chi_{ij} = -\lb -a_0  +   a_1 u_1  +  a_2 u_2 +  a_3  u_3 \rb ,\label{eq:chi_ij_calculate}
	\ee
	where
	\bea
	&&u_1= \frac{1}{4}\Tr \lb \tau_3 \sigma_1 \tau_i\sigma_j \tau_3 \sigma_1\tau_i\sigma_j\rb,\nonumber\\
	&&u_2= \frac{1}{4}\Tr \lb 1_2 \sigma_2 \tau_i\sigma_j 1_2 \sigma_2 \tau_i\sigma_j\rb,\nonumber\\
	&&u_3= \frac{1}{4}\Tr \lb 1_2 \sigma_3 \tau_i\sigma_j 1_2 \sigma_3 \tau_i\sigma_j\rb.\nonumber
	\eea
	We note that all $a_0, a_1, a_2, a_3$ are of a positive sign.
	The signs of different terms in Eq.\eqref{eq:chi_ij_calculate} are such, 
	that $\chi_{ij}$ takes maximal value if and only if $u_0=u_2=u_3=-1$, which leads to two solutions $i=1,j=1$ and $i=2,j=1$. Therefore, the following two channels have the strongest Stoner instability:
	\be
	O = \tau_1 \sigma_1, \quad \tau_2\sigma_1. \label{eq:mother state order_S}
	\ee
	Being off-diagonal in valleys, these states describe isospin-polarized phases with valley coherence spontaneously generated as a result of Stoner's instability. 
	To assess whether these two candidate orders can play a role of 
	mother states for superconductivity below we consider isospin fluctuations for these two states.

	\subsection{Pairing interaction mediated by the soft isospin fluctuation modes}
	As in the main text, we use $\phi$ to represent the collective mode that undergoes softening near the boundary of a partially isospin ordered (PIP) phase. The soft mode can be described by a Hamiltonian of the following form
	\be
	H_s = \frac{1}{2}|\p_t \phi|^2+ \frac{K}{2}|\p_x \phi|^2 + \frac{J}{2}|\phi|^2  + \frac{U}{2}|\phi|^4,\label{eq:H_s_S}
	\ee
	where $U$ is always positive, $J>0$ outside the 
	PIP phase, $J<0$ inside 
	the PIP phase. As the field $\phi$ describes the fluctuation of the PIP order parameter, it transforms under the space group in the same way as $\Psi^\dagger\tau_1\sigma_1\Psi$. The soft-mode Hamiltonian Eq.\eqref{eq:H_s_S} respects the full symmetry of the system since both $|\phi|^2$ and $|\p_x \phi|^2$ are invariant under the space group. 
	From this Hamiltonian, the Green's function for soft mode in disordered (normal) phase is
	\be
	D_{\phi\phi}(i\omega,p) = -\langle \phi(i\omega,p) \phi(-i\omega,-p)\rangle = -\frac{2}{\omega^2+Kp^2+J}.
	\ee
	We use the following Hamiltonian to represent the coupling between soft modes and electrons:
	\be
	H_{e-s} = \gamma \sum_{\vec p, \vec q}\phi_{\vec q} \Psi^\dagger_{\vec p+\vec q} \tau_1 \sigma_1 \Psi_{\vec p} +h.c., 
	\ee
	Here, we note that the field $\phi$ in this model carries momentum $2K$ since the Dirac bilinear $\Psi^\dagger \tau_1 \sigma_1 \Psi$ carries momentum $2K$. This soft-mode-electron coupling Hamiltonian also respects the full symmetry of the system. 
	Taking the PIP order of the form $\tau_1 \sigma_1$, and using the perturbation theory at the second order of $\gamma$, we obtain a soft-mode-mediated electron-electron interaction \addZ{represented by the following action: 
	\be
	A_{ee} = \sum_{\omega,\nu,\nu', \vec p, \vec k,\vec k'}\tilde{V}_{i\omega,p} \overline{\Psi}_{i\nu'+i\omega,\vec k'+\vec p} \tau_1 \sigma_1 \Psi_{i\nu',\vec k'}\overline{\Psi}_{i\nu-i\omega,\vec k-\vec p} \tau_1 \sigma_1 \Psi_{i\nu,\vec k} \label{eq:H_ee_S}
	\ee}
	where \addZ{$\Psi_{i\nu,\vec k}$ and $\overline{\Psi}_{i\nu,\vec k}$ are four-component Fermionic fields describing four-component Dirac electrons.}
	Outside the $\tau_1 \sigma_1$ phase, $\tilde{V}(i\omega,p)$ takes the form
	\be
	\tilde{V}_{i\omega,p} = -\gamma^2 (-D_{\phi\phi}(i\omega,\vec p )) =	-\frac{2\gamma^2}{\omega^2+\Omega_p^2}, 
	\ee
	where $\Omega_p =  \sqrt{K p^2 +J}$. 
	
	The soft-mode-mediated interaction inside 
	PIP phase takes on a different form. 
	Inside $\tau_1 \sigma_1$ phase, isospin polarization has a nonzero expectation value $\phi_0 = \langle\Psi^{\dagger}\tau_1 \sigma_1\Psi\rangle \neq 0$. Since we are interested in the fluctuations of the order parameter, we define $\phi = \phi_0+\delta\phi$ where $\delta\phi = \delta \phi_{\parallel} + i\delta\phi_{\perp}$, where parallel and perpendicular components correspond to amplitude and phase fluctuations. Then, we rewrite  Eq.\eqref{eq:H_s_S} in terms of $\delta\phi$: 
	\bea
	&&H_s[\phi] = H_s[\phi_0] + H_s[\delta \phi(x,t)], \\
	&& H_s[\phi_0] =   \frac{J}{2} |\phi_0|^2  + \frac{U}{2}|\phi_0|^4,\\
	&&H_s[\delta \phi(x,t)] = \frac{1}{2}|\p_t \delta\phi|^2+ \frac{K}{2}|\p_x \delta \phi|^2 + \lp \frac{J}{2} + 3U|\phi_0|^2\rp \delta\phi_{\parallel}^2  + \frac{U}{2}\phi_{\parallel}^4.\label{eq:H_s_2S}
	\eea
	By minimizing $H_s[\phi_0]$, we find $|\phi_0|^2=-\frac{J}{2U}$ (which is positive since $J<0$). Using this result, Eq.\eqref{eq:H_s_2S} is simplified to:
	\be
	H_s[\delta \phi(x,t)] = \frac{1}{2}|\p_t \delta\phi|^2+ \frac{K}{2}|\p_x \delta \phi|^2 - J \delta\phi_{\parallel}^2  + \frac{U}{2}\delta \phi_{\parallel}^4.
	\ee
	The resulting propagator of soft fluctuations is 
	\be
	D_{\delta\phi\delta\phi}(i\omega,\vec p) = -\langle \delta \phi_{\parallel}(i\omega,\vec p)^2 + \delta \phi_{\perp}(i\omega,\vec p)^2 \rangle = -\frac{1}{\omega^2+Kp^2-2J} -\frac{1}{\omega^2+Kp^2}.
	\ee
	The coupling between the soft mode and electrons can 
	be written, accordingly, as a sum of two parts:
	\be
	H_{e-s} = \gamma\phi_0 \sum_{\vec p}\Psi_{\vec p}^\dagger \tau_1 \sigma_1 \Psi_{\vec p}  + \gamma \sum_{\vec p, \vec q}\delta\phi_{\vec q} \Psi_{\vec p+\vec q}^\dagger \tau_1 \sigma_1 \Psi_{\vec p}  +{\rm h.c.}. 
	\ee
where $\delta \phi_\vec q = \int dx \delta \phi(\vec x) e^{-i\vec q\cdot \vec x} $The first term of this expression can be absorbed into the single-particle qHamiltonian, whereas the second term generates a soft-mode-mediated interaction between electrons. The resulting effective e-e interaction is of a form similar to Eq.\eqref{eq:H_ee_S}; the expression of $\tilde{V}$ inside ordered phase is slightly different from the one in the disordered phase. We find:
	\be
	\tilde{V}_{i\omega,\vec p} = -\gamma^2 (-D_{\delta\phi\delta\phi}(i\omega,\vec p )) =	- \frac{\gamma^2}{\omega^2+Kp^2-2J} -\frac{\gamma^2}{\omega^2+Kp^2}, 
	\ee
	In summary, putting the two cases together, the effective e-e interaction mediated by fluctuations 
	takes the form of Eq.\eqref{eq:H_ee_S}, 
	with $\tilde{V}$ in the two cases given by 
	\bea
	&&\tilde{V}_{i\omega,p}=  
	\begin{cases}
		\frac{\gamma^2}{-\omega^2-\omega_{p\parallel}^2}+\frac{\gamma^2}{-\omega^2- \omega_{p\perp}^2}, \quad\text{inside $\tau_1 \sigma_1$ phase},\\
		\frac{2\gamma^2}{-\omega^2-\Omega_p^2}, \quad\text{outside $\tau_1 \sigma_1$ phase}.
	\end{cases}
	\eea
	where
	\be
	\Omega_p = \sqrt{K p^2 + J}, \quad \omega_{p\perp} = \sqrt{K} p, \quad \omega_{p\parallel} = \sqrt{K p^2 - 2J}.
	\ee
	In the main text we exploit this interaction as a pairing glue for superconductivity. Softening of the polarization mode $\phi$ near the PIP phase boundary leads to a stronger interaction at the PIP onset, which explains that in the measured phase diagram superconductivity occurs at the PIP phase boundary.
	
	\subsection{Identifying the strongest pairing channel}
	As we have shown in the main text, the linearized pairing gap equation in an arbitrary pairing channel $\langle\Psi^T\tilde{O}\Psi\rangle$ takes the following form:
	\be
	- \int \frac{d\omega}{2\pi} \frac{d^2p}{\lp 2\pi \rp ^2} \tilde{V}  O^T G^T(-i\omega,-p)\tilde{O} G(i\omega,p) O = \tilde{O}, \label{eq:pairing_instability_RTG_S}
	\ee
	where $O=\tau_1\sigma_1$ is the PIP order parameter.
	Here, the quantity $\tilde{O}$ can be an arbitrary complex-valued $4\times4$ matrix. We note that we have ignored the frequency and momentum dependence of $\tilde{V}$ for simplicity.

	We note that the matrix structure in Eq.\eqref{eq:pairing_instability_RTG_S} resembles Eq.\eqref{eq:Stoner_matrix_1S}. 
	Analyzing Eq.\eqref{eq:pairing_instability_RTG_S} in a same way as in the analysis of Eq.\eqref{eq:Stoner_matrix_1S} above, we see 
	that the problem of strongest pairing channel is equivalent to maximizing the following function:
	\be
	\tilde{F}[\tilde{O}] = -\int \frac{d\omega d^2p}{\lp 2\pi \rp ^3}  \frac{1}{4}\Tr \lb \tilde{O}^\dagger O^T G^T(-i\omega,-p) \tilde{O} G(i\omega,p) O\rb \label{eq:tildeF_S}
	\ee
	under the normalization constraint $\frac{1}{4}\Tr \tilde{O}^\dagger \tilde{O} =1$.
	
	Plugging the propagators $G$ and $G^T$ in Eq.\ref{eq:tildeF_S}, and carrying out integration over $\omega$ and $p$ yields
	\begin{align}
	\tilde{F}[O] = \tilde{a}_0\tilde{u}_0 - \tilde{a}_1\tilde{u}_1  - \tilde{a}_2\tilde{u}_2  + \tilde{a}_3\tilde{u}_3 + \tilde{b} \tilde{v}   \label{eq:maximize_tildeF}
	\end{align}
	Here $\tilde{V}<0$, the quantities $\tilde{u}_0$, $\tilde{u}_1$, $\tilde{u}_2$, $\tilde{u}_3$ and $\tilde{v}$ are quadratic functions of $\tilde{O}$,
	and the coefficients $\tilde{a}_i$'s and  $\tilde{b}$ are defined as follows:
	\begin{align}
	&\tilde{a}_0 = \int \frac{d^2p}{\lp 2\pi \rp ^2} \tilde{A}(E),\ \ 
	\tilde{a}_1 = \int \frac{d^2p}{\lp 2\pi \rp ^2}\alpha^2 \lp p_1^3-3p_1p_2^2\rp^2 \tilde{B}(E),\nonumber
	\\
	&\tilde{a}_2 = \int \frac{d^2p}{\lp 2\pi \rp ^2}\beta^2(3p_1^2p_2-p_2^3)^2 \tilde{B}(E), \nonumber
	\\
	&\tilde{a}_3 = \int \frac{d^2p}{\lp 2\pi \rp ^2}D^2 \tilde{B}(E),\quad 
	\tilde{b} =  \int \frac{d^2p}{\lp 2\pi \rp ^2} \mu D \tilde{B}(E),\nonumber
	\end{align}
	where we have introduced notations $\tilde{A}(E) =   \frac{1}{4}( \frac{E\Theta(E-|\mu|)}{E^2-\mu^2} + \frac{|\mu|\Theta(|\mu|-E)}{\mu^2-E^2})$ and $\tilde{B}(E) =  \frac{1}{4}( \frac{\Theta(E-|\mu|)}{E(E^2-\mu^2)} + \frac{\Theta(|\mu|-E)}{|\mu|(\mu^2-E^2)})$.
	The quantities $\tilde{u}_i$'s and $\tilde{v}$, which describe possible superconducting pairing order types, are defined as
	\bea
	&&\tilde{u}_0 =  \frac{1}{4}\Tr [\tilde{O}^\dagger\tau_1\sigma_1 \tilde{O}\tau_1\sigma_1],  \quad
	\tilde{u}_1 =  \frac{1}{4}\Tr [\tilde{O}^\dagger \tau_2 1 \tilde{O}\tau_2 1],  \nonumber\\
	&&\tilde{u}_2 =  -\frac{1}{4}\Tr [\tilde{O}^\dagger \tau_1\sigma_3\tilde{O}\tau_1 \sigma_3], \quad
	\tilde{u}_3 =  \frac{1}{4}\Tr [\tilde{O}^\dagger\tau_1 \sigma_2 \tilde{O}\tau_1 \sigma_2],  \nonumber\\
	&&\tilde{v} =  \frac{1}{4}\Tr [\tilde{O}^\dagger\tau_1 \sigma_1 \{\tilde{O},1\sigma_3 \} \tau_1 \sigma_1], \label{eq:tilde v_S}
	\eea
	where $\{\tilde{O},1\sigma_3 \} = \tilde{O}1\sigma_3 + 1\sigma_3\tilde{O}$. 
	
	As a reminder, $\tilde{O}$ can be an arbitrary complex-valued $4\times 4$ matrix, so we express it under Pauli matrix basis $\tilde{O} = \tilde{d}_{ij}\tau_{i}\sigma_{j}$ ($i,j = 0,1,2,3$) where $\tilde{d}_{ij}$'s are complex numbers. 
	With this, we 
	see that $\tilde{u}_0$, $\tilde{u}_1$, $\tilde{u}_2$ and $\tilde{u}_3$ generates diagonal terms $|d_{ij}|^2$, whereas $\tilde{v}$ generates off-diagonal terms that hybridize the $\tau_i 1$ and $\tau_i \sigma_3$ components of $\tilde{O}$ 
	, i.e. $d_{i3}^*d_{i0}$ and $d_{i0}^*d_{i3}$. As a result, the candidate channels that maximize the function $\tilde{F}[\tilde{O}]$ should take one of the following forms
	\be
	\tilde{O} = \tau_i \sigma_1, \tau_i\sigma_2, \tau_i (\xi 1 + \eta \sigma_3)
	\ee
	where $\xi$ and $\eta$ are complex numbers. 
	
	Plugging these three candidate 
	 forms into $\tilde{F}[\tilde{O}]$, after a straightforward calculation we find
	\begin{align}
	&&\max(\tilde{F}[\tau_i\sigma_1]) = \tilde{F}[\tau_1\sigma_1] = \tilde{a}_0  + \tilde{a}_1  - \tilde{a}_2- \tilde{a}_3 \nonumber \\
	&&\max(\tilde{F}[\tau_i\sigma_2]) = \tilde{F}[\tau_3\sigma_2] = \tilde{a}_0  + \tilde{a}_1  + \tilde{a}_2- \tilde{a}_3 \label{eq:F_pairing_sigma_1,2}
	\end{align}
	whereas for $\tau_i (\xi 1 + \eta \sigma_3)$ channel, $\tilde{F}[\tilde{O}]$ is maximized at
	\be
	\tilde{O} = \tau_1 (\xi_* 1 +\eta_* \sigma_3), 
	\label{eq:strongest_pairing_channel_S}
	\ee
	The maximal value of $\tilde{F}$ is
	\be
	\tilde{F}[\tau_1 (\xi_* 1 + \eta_* \sigma_3)] = \tilde{a}_1 +\tilde{a}_2 + \sqrt{(\tilde{a}_0 +\tilde{a}_3)^2 - 4\tilde{b}^2} \label{eq:F_pairing_sigma_0,3}
	\ee
	where $\xi_* = \frac{-\tilde{a}_0-\tilde{a}_3}{\sqrt{(\tilde{a}_0 +\tilde{a}_3)^2 - 4\tilde{b}^2}}$, $\eta_* = \frac{2\tilde{b}}{\sqrt{(\tilde{a}_0 +\tilde{a}_3)^2 - 4\tilde{b}^2}}$. 
	Using the expression of $\tilde{a}_0$,$\tilde{a}_3$ and
	$\tilde{b}$, it is not hard to show that $\sqrt{(\tilde{a}_0 +\tilde{a}_3)^2 - 4\tilde{b}^2} > \tilde{a}_0 -\tilde{a}_3$. Therefore, comparing Eq.\eqref{eq:F_pairing_sigma_1,2} and Eq.\eqref{eq:F_pairing_sigma_0,3}, we conclude that the strongest pairing channel is the one given in Eq.\eqref{eq:strongest_pairing_channel_S}

	Lastly, we discuss the case when PIP order is $\tau_2\sigma_1$. In this case, one can repeat the same procedure. The form of Eq.\eqref{eq:tildeF_S} will be unchanged, while Eq.\eqref{eq:tilde v_S} should be rewritten using the following substitution:
	\be
	\tilde{O}^\dagger \tau_1 \tilde{O} \tau_1 \rightarrow -\tilde{O}^\dagger\tau_2 \tilde{O} \tau_2,\quad \tilde{O}^\dagger\tau_2 \tilde{O} \tau_2 \rightarrow -\tilde{O}^\dagger \tau_1 \tilde{O} \tau_1
	\ee
	Under this substitution, the values of $\tilde{F}$ for $\tau_1 \sigma_j$ and $\tau_2 \sigma_j$ channels ($j=0,1,2,3$) are invariant, while the values of $\tilde{F}$ in $\tau_3 \sigma_j$ and $\tau_0 \sigma_j$ channels pick up a minus sign as compared to their values in the case when the PIP order is $\tau_1\sigma_1$. With this in mind, it is straightforward to write down the value of $\tilde{F}$ in all channels. 
	Ultimately, we find that the maximal values of $\tilde{F}$ in each of the three types of channels $\tau_i \sigma_1$, $\tau_i \sigma_2$ and $\tau_i (\xi 1 + \eta \sigma_3)$ are the same as Eq.\eqref{eq:F_pairing_sigma_1,2} and Eq.\eqref{eq:F_pairing_sigma_0,3}. 
	Therefore, no matter which of the two candidate PIP orders $\tau_1\sigma_1$ and $\tau_2\sigma_1$ is chosen as the actual PIP order, the strongest pairing instability always occurs in the pairing channel given in Eq.\eqref{eq:strongest_pairing_channel_S}. 

\end{widetext}

\end{document}